\documentclass[pra,twocolumn,showpacs,superscriptaddress]{revtex4}
\usepackage{amsmath}
\usepackage{amsfonts}
\usepackage{graphicx} 
\usepackage{bm}
\usepackage{enumerate} 
\usepackage{color}
\usepackage{hyperref}

\begin{document}
\title{Uniform cross phase modulation for nonclassical radiation pulses}
\author{Karl-Peter Marzlin}
\affiliation{Department of Physics, St.~Francis Xavier University, Antigonish, Nova Scotia, B2G 2W5, Canada}
\affiliation{Institute for Quantum Information Science, University of Calgary, Calgary, Alberta T2N 1N4, Canada}
\author{Zeng-Bin Wang}
\affiliation{Institute for Quantum Information Science, University
of Calgary, Calgary, Alberta T2N 1N4, Canada}
\author{Sergey A.~Moiseev}
\affiliation{Institute for Quantum Information Science, University
of Calgary, Calgary, Alberta T2N 1N4, Canada}
\affiliation{Kazan Physical-Technical Institute of the Russian Academy
 of Sciences, 10/7 Sibirsky Trakt, Kazan, 420029, Russia}

\author{Barry C.~Sanders}
\affiliation{Institute for Quantum Information Science, University
of Calgary, Calgary, Alberta T2N 1N4, Canada}
\date{\today}
\begin{abstract}

We propose a scheme to achieve a uniform cross phase modulation (XPM)
for two nonclassical light pulses and study its application 
for quantum non-demolition measurements of 
the photon number in a pulse and for controlled phase gates
in quantum information. We analyze the scheme by quantizing a common
phenomenological model for classical XPM.
Our analysis first treats the ideal case of equal cross-phase modulation and 
pure unitary dynamics. This establishes the groundwork for more complicated studies 
of non-unitary dynamics and difference in phase shifts between the two pulses
where decohering effects severely affect the performance of the scheme.
\end{abstract}

\maketitle

\section{Introduction}
Optical cross phase modulation (XPM) is a specific variant
of nonlinear optical phenomena of Kerr type in which
the refractive index $n_1$ of light in pulse 1 varies linearly
with the intensity $I_2$ of another light field, so that
$\epsilon_1 = 1+ \chi_1 + \chi_3 I_2$, with $ \epsilon_1 = n_1^2$
the permittivity of the medium, $\chi_1$ the linear, and $ \chi_3$ the 
nonlinear XPM susceptibility, respectively. In ordinary optical
media this effect is small and requires large intensities, but
the proposal by Schmidt and Imamo{\v g}lu \cite{Schm96} to
generate giant nonlinearities using electromagnetically induced
transparency (EIT)
\cite{Harr97,boller91,Lukin97} has made very large Kerr coefficients possible
\cite{Schm96,Harr99,Hauu99,Bajcsy03,kang03} and may even lead to nonlinear effects
at the single-photon level \cite{Luki00,Pett02,Mats03,ottaviani03,Pett04,rebic04,Zeng06}.

XPM is a strong candidate for the design of 
optical quantum controlled-phase gates (CPG)
for photonic quantum information processing~\cite{Chuu95,nemoto05,munro05},
in which two single-photon pulses would become entangled.
For sufficiently high fidelity, XPM would allow
the construction of deterministic gates, as opposed
to the non-deterministic optical CPG that are  based on linear optics
\cite{Sana95} and establish entanglement through
measurements~\cite{Knil01}.
It has been suggested that
double EIT -- EIT for both pulses with matched group
velocities -- would generate the maximal XPM phase
shift because matched group velocities for both photons
would maximize their interaction time \cite{Luki00,Pett02,Mats03,ottaviani03,Pett04,rebic04,Zeng06}. 
However, XPM based on double EIT still faces some challenges:
(i) to achieve sufficiently high intensities at the two-photon
level, the photon pulses must be tightly confined in the transversal
direction; (ii) if the nonlinear medium has a finite response time,
the matter-light interaction unavoidably induces noise. It was shown 
by Shapiro \cite{Shap06} that for a large class of XPM models
this would limit the fidelity of a CPG to only about 65\%.
Further theoretical studies on this topic confirmed these results \cite{shapiro:022301,shapiro:njp2007,koshino:063807,leungXPM2008}.
(iii) For matched co-propagating pulses, each point of
the pulse will experience a different XPM phase shift because the
intensity, and hence the refractive index $n_1$, varies over the
shape of the pulse. This would severely affect the entanglement
between the pulses.

In this paper we will address the third
problem and do not deal with problem (i) and (ii), i.e., we will
assume that transverse confinement of the photon pulses can be
achieved by some means, such as hollow core fibers \cite{PhysRevLett.75.3253,Gehring:JLightwTechn2008,hollowFibres,londero:043602}
or nano wires \cite{LukinNanoWire}, and that the medium's response
time is so short that the polarizability of the medium 
reacts instantaneously to a photon's electric field.
The omission of the noise associated with a finite response time
is also made for clarity because, despite that we are working in the instantaneous regime,
we will show that very similar effects may appear if the two light pulses have different group
velocities.
\begin{figure}[t]
\includegraphics[width=8cm]{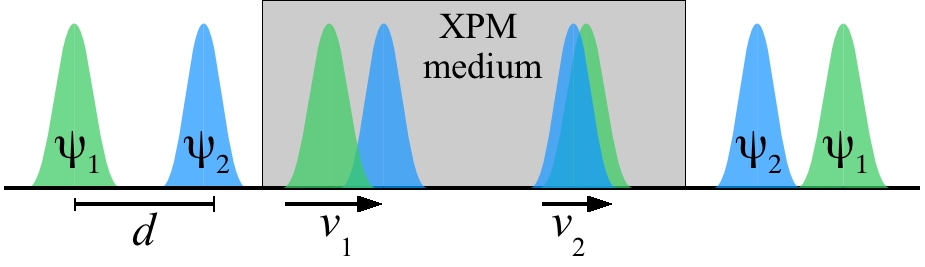}
\caption{\label{initModes} 
Scheme to achieve uniform XPM: two light pulses $\psi_1, \psi_2$
travel at different group velocities $v_1 >v_2$ through an XPM medium such
that $\psi_1$ can overtake $\psi_2$.
$d$ denotes the initial distance between the two pulses.
}
\end{figure}
To solve problem (iii) we extend an idea of Rothenberg \cite{Josh93}  who demonstrated a
uniform XPM phase shift for classical pulses in birefringent fibers:
two light pulses travel through a XPM medium
with different group velocities $v_1>v_2$, see Fig.~\ref{initModes}. 
Pulse 2 (blue) reaches
the Kerr medium first but pulse 1 (green) is faster and leaves it
first. While the pulses overlap they interact via the Kerr effect, which
is proportional to the pulse intensity. Because pulse 1 overtakes 
pulse 2 the acquired phase shift will be averaged
over the pulse shape and thus be nearly uniform.

Here we generalize this idea to characterize the propagation of 
quantized light pulses and show that a uniform phase shift can 
also be achieved for quantum interference effects.  
We will show that quantizing the classical XPM equations
is generally a subtle problem which may make the introduction
of noise terms similar to those in problem (i) necessary.
We exploit our results to suggest experiments for 
quantum non-demolition (QND) measurement
of photon numbers and CPG. 

This paper is organized as follows. In Sec.~\ref{Sec:pulseProp} we will
present a phenomenological quantum model for XPM between two pulses with 
different group velocities and discuss the two cases of unitary and non-unitary
dynamics in this model. In Sec.~\ref{Sec:XPMunitary} we present proposals
for QND measurement of photon numbers and CPG and show that they can have
a high fidelity in this case. The analysis of these proposals for the non-unitary case 
is presented in Sec.~\ref{sec:nonUnitary}. Several appendices contain details
of our derivation.

\section{Pulse propagation with mismatched group velocities}\label{Sec:pulseProp}
We consider two quantized light pulses that travel through an XPM medium
with different group velocities $v_1>v_2$. We assume that the light pulses
are transversally confined so that we can restrict the model to one spatial
dimension. In classical fiber optics this situation is commonly described by the
set of equations \cite{agrawal06,PTL6-733,JRLR29-57}
\begin{align}
  \left (\frac{1}{v_{i}}\partial_t+\partial_z \right ) {\cal E}_i(z,t)&=
-{\rm i} \gamma_i (I_{i}(z,t) + 2 I_{3-i}(z,t)) {\cal E}_i(z,t) ,
\label{xpmEqnClass} \end{align}
with imaginary unit ${\rm i}=\sqrt{-1}$.
Here $ {\cal E}_i$ is the field amplitude of pulse $i=1,2$, 
$I_i=|{\cal E}_i|^2$ is proportional to the intensity, and $\gamma_i$
is the self phase modulation (SPM) coefficient.
The XPM coefficient is given by $2\gamma_i$.
This model is local and instantaneous; i.e.,
the light field ${\cal E}_1(z,t)$ is only affected by ${\cal E}_2(z,t)$
and vice versa.

In the context of SPM, Joneckis and Shapiro 
have shown that in a local and instantaneous quantum model, where
$ {\cal E}_i$ represents a field operator, 
infinite zero-point fluctuations do occur \cite{shapiro:josab1993}.
They suggested
taking a finite response time of the medium into account to avoid
singularities. 
We instead propose a model in which the medium's
response is still instantaneous but spatially nonlocal, i.e., 
${\cal E}_1(z,t)$ can be affected by ${\cal E}_2(z',t)$.
The most general macroscopic model that describes such an instantaneous
interaction between two light pulses of different group velocities
is then given by
\begin{align}
  \left (\frac{1}{v_{i}}\partial_t+\partial_z \right ) {\cal E}_i(z,t)&=
-{\rm i} \int_{-\infty}^{\infty} dz'I_{3-i}(z',t)
\nonumber \\ &\times 
  V_i(z-z'){\cal E}_i(z,t) .
\label{xpmEqn} \end{align}
Here and henceforth  ${\cal E}_i$ denotes the field operator and
$I_i(z,t)={\cal E}_i^{\dagger}(z',t){\cal E}_i(z',t)$ 
the intensity operator.
The quantities $V_i(z-z')$ are spatially nonlocal interaction potentials
between the two pulses generated by the atomic medium.
To avoid causality violation its support must be much
smaller than the wavelength of light. We will assume that 
$V_i(z-z')$ is essentially zero if the distance between $z$ and
$z'$ is much larger than the Bohr radius. 
One can consider the introduction of nonlocal potentials as a regularization
of the classical theory (\ref{xpmEqnClass}) by smearing out the point interaction.
In the limit of sharp potentials,
$V_i(z) = {\cal V}_i \delta(z)$, the regularized model reduces to Eq.~(\ref{xpmEqnClass})
for XPM coefficient $\gamma_i = {\cal V}_i/2$. 

Multi-photon pulse dynamics should account for SPM, but we omit this effect here in order to focus on XPM. 
When XPM is placed in the context of interferometry, which gives phase an operational meaning, pairs of nonlinear 
media in both paths can offset SPM \cite{Sanders:JOSAB92}. Alternatively SPM is not present for single-photon 
pulses because a single photon cannot induce SPM on itself. 

The multi-mode
field operators $\mathcal{E}_i (z)$ satisfy the commutation relations 
\begin{align}
  [ {\cal E}_i(z,t),{\cal E}_j^{\dagger}(z',t)]&=\frac{\hbar
  \omega_i}{2 \epsilon_0 A} \delta_{ij} \delta(z-z')
  \equiv \eta \delta_{ij}  \delta(z-z') ,
\end{align}
with $\omega_i$ the central frequency of pulse $i$ and $A$
the transverse area of the pulses. 
The field is assumed to be in a single mode with the $i^{\rm th}$ longitudinal mode 
function given by $\psi_i(z)$.
Photons in this mode are annihilated by the operator
\begin{align}
a_i[u] &=\frac{1}{\sqrt{\eta}}\int_{-\infty}^{\infty} dz\,
\psi^{*}_i(z-u)\, \hat {\cal{E}}_i(z,0) \; ,
\label{eq:aModeOp}\end{align}
For later use we have included a possible shift $u$ of the
wavefunction. If the shift is zero we will sometimes suppress this
notation and write $a_i \equiv a_i[0]$.

If Eq.~(\ref{xpmEqn}) is taken to be the dynamical equation of the quantum
fields ${\cal E}_i$ one has to be careful with its interpretation.
It can be shown that Eq.~(\ref{xpmEqn}) can be expressed as
$\partial_t {\cal E}_i = -{\rm i} [{\cal E}_i, H]$ for some Hamiltonian $H$
if and only if $v_1 V_1(z)= v_2 V_2(z)$. In other words, a unitary evolution
of the quantum fields will only occur if the two interaction potentials
are proportional to each other, with the proportionality factor given by
the ratio of the group velocities. In absence of absorption or other
decohering processes one therefore would assume that in the quantum
model this relation must be fulfilled. However, in the corresponding
classical models for XPM pulse propagation this is
generally not the case \cite{agrawal06,PTL6-733,JRLR29-57}.

In our subsequent analysis we will therefore
first assume that the evolution is unitary and analyze
our proposals within this framework.
In Sec.~\ref{sec:nonUnitary}  we then will develop a
consistent phenomenological quantum model for non-unitary evolution
using techniques of open quantum systems and reassess the proposals
for this case \cite{footnote1}.

\section{XPM for unitary evolution}\label{Sec:XPMunitary}

To devise a scheme for CPG and QND measurements of the photon number
we will assume
that the dynamics is unitary; in this case the interaction potentials can generally be
written as  $V_i(z) = V(z)/v_i$, with $V(z)$ a Hamiltonian interaction potential.
However, we will make use of this relation only after deriving general
results in order to facilitate the extension to non-unitary dynamics.

The solution of Eq.~(\ref{xpmEqn}) for the propagation of light in an
infinitely extended XPM medium is
\begin{align}
{\cal E}_i(z,t)&= U_{3-i}(z,t) {\cal E}_i(z-v_{i}t,0),
\label{solu}
\\
U_{i}(z,t)&=\exp \Bigg (-{\rm i}\int_{-\infty}^{\infty} \hat
I_{i} (z',0) \frac{ v_{3-i}}{v_{i}-v_{3-i}}
\nonumber \\ & \hspace{5mm} \times 
\theta_{3-i} \left (z-z' -v_{3-i} t \, ,\, [v_{i}-v_{3-i}]t \right ) dz' \Bigg ) ,
\label{eq:U1}\\
  \theta_i (z_1,z_2) &\equiv
  \int_0^{z_2} dz' \, V_i(z_1-z')
\; .
\end{align}
The function $\theta_i(z_1,z_2)$ will play a central role in determining the Kerr
phase shift. One important property is that for a symmetric potential, $V_i(-z)=V_i(z)$,
we have $ \theta_i (-z_1,-z_2) =  -\theta_i (z_1,z_2)  $.

\subsection{Quantum non-demolition measurement of photon numbers}\label{sec:qndUnitary}
As a first application we consider a
QND measurement of the number of photons \cite{San89}
in mode 2. This can be accomplished by sending a strong classical
pulse in mode 1, which can be described by a coherent state
$ |\alpha \rangle $,
together with an $n$ photon pulse $|n \rangle $ in mode 2 through
the XPM medium. Single-mode treatments predict that in this case
the phase of the classical pulse will be shifted by an amount
that is proportional to $n$. 
Here we show that the uniform cross phase shift that we suggest will accomplish precisely this.

The initial state of the two light pulses before they start
to interact takes the form $ |\psi \rangle = |\alpha \rangle \otimes
|n \rangle $ where the two kets refer to mode 1 and 2, repectively.
Using the shift operator $D_1 (\alpha) = \exp (\alpha a_1^\dagger
-\alpha^* a_1)$ this can be expressed as
\begin{equation} 
   |\psi \rangle = \frac{ 1}{\sqrt{n!}} D_1 (\alpha) (a_2^\dagger)^n
   |0 \rangle   .
\end{equation} 
The complex amplitude of the classical pulse at time $t$ and position
$z$ is given by $\langle \psi | {\cal E}_1 (z,t) | \psi \rangle $.
Its phase can be measured using homodyne detection \cite{tyc:JPA2004}, for instance,
which more specifically measures the observable 
$ X(\vartheta)= e^{{\rm i}\vartheta} {\cal E}_1 (z,t) + e^{-{\rm i}\vartheta} {\cal E}_1^\dagger  (z,t)$.
Using Eq.~(\ref{eq:comm1}) it is easy to see that
\begin{equation} 
  D_1^\dagger  (\alpha){\cal E}_1 (z,0) D_1 (\alpha) =
  {\cal E}_1 (z,0) + \alpha \sqrt{\eta} \psi_1(z) \; .
\end{equation} 
Exploiting this and solution (\ref{solu}) we get
\begin{align}
  \langle \psi | {\cal E}_1 (z,t) | \psi \rangle 
  &=
   \frac{ \alpha \sqrt{\eta}}{n!}  \psi_1(z-v_1 t)
   \langle 0| a_2^n U_2(z,t)  (a_2^\dagger)^n |0 \rangle 
\nonumber \\ &=  \frac{ \alpha \sqrt{\eta}}{n!}  \psi_1(z-v_1 t) 
  \langle 0| a_2^n  (
  \tilde{a}_2^\dagger[0;z,t]  )^n |0 \rangle ,
\label{eq:QNDresult1}\end{align}
where we have used Eq.~(\ref{eq:unitransA}) and introduced
an XPM-modified annihilation operator
\begin{align}
    \tilde{a}_i[u; z,t] &\equiv 
   \frac{1}{\sqrt{\eta}}\int_{-\infty}^{\infty} dz'\,
  \psi^{*}_i(z'-u)\,   \hat {\cal{E}}_i(z',0) 
\nonumber \\ &\times
  e^{ 
   -{\rm i}\eta \frac{ v_{3-i}}{v_i-v_{3-i}} 
   \theta_{3-i}(z-z' -v_{3-i} t, (v_i-v_{3-i} )t)
  } \; .
\label{eq:aTilde}\end{align}
The physical interpretation of Eq.~(\ref{eq:aTilde}) is that the
wavepacket $ \psi^{*}_i(z'-u)$ is multiplied by a spatially varying
phase factor that is given by the exponential in Eq.~(\ref{eq:aTilde})
and incorporates the effect of XPM on the light pulses.

The expectation value for field 2 in Eq.~(\ref{eq:QNDresult1}) can then be reduced to 
\begin{align}
   \langle 0| a_2^n  (  \tilde{a}_2^\dagger[0;z,t] 
  )^n |0 \rangle  &=
  n!  [a_2 ,  \tilde{a}_2^\dagger[0;z,t]\, ]^n,
\end{align}
so that
\begin{align}
  \langle {\cal E}_1 (z,t) \rangle 
  &=  \alpha \sqrt{\eta}  \psi_1(z-v_1 t)
\nonumber \\ & \times 
  \bigg (
   \int dz'\,  |\psi_2(z')|^2   
 e^{ 
   {\rm i}\eta \frac{ v_{1}}{\Delta v} 
   \theta_1(z'-z +v_{1} t, \Delta v\, t)} \bigg )^n,
\label{eq:QNDresult}\end{align}
with $\Delta v \equiv v_1 - v_2$.

To better understand the physical implications of
Eq.~(\ref{eq:QNDresult}) we consider the specific configuration
depicted in Fig.~\ref{initModes}. The classical mode $\psi_1$ is initially
centered around the origin while the $n$-photon pulse in mode $\psi_2$
is a distance $d$ to the right of it \cite{footnote2}.
Both pulses are moving to the right, but pulse 1 is faster.
The first line in Eq.~(\ref{eq:QNDresult}) 
is basically the amplitude of the classical pulse $\psi_1$ at time $t$ in absence of the
XPM medium. The pulse is centered around $z=v_1
t$. Hence, to achieve an maximum phase contrast, we should observe the
field at this point.  The exponential in Eq.~(\ref{eq:QNDresult}) is
then proportional to $  \theta_1(z', \Delta v\, t)$.

In Appendix \ref{app:phaseShift} it is shown that for any potential $V_i(z)$ that
is consistent with causality, the function $  \theta_1(z', \Delta v\, t)$
is nearly constant between the lines $\Delta v \, t = z'$ and $z'=0$ and zero outside
of this range. The support of the initial wavepacket $\psi_2(z')$ is in the
area $z'>0$ and peaked around $z'=d$. For sufficiently large times, such that $\Delta v \, t  \gg d$, the 
support of $\psi_2(z')$ is therefore completely inside the domain where 
$\theta_1 (z', \Delta v t) \approx {\cal V}_1 $, with the constant
\begin{equation} 
  {\cal V}_i \equiv \int_{-\infty}^\infty dz\, V_i(z) .
\label{calViDef}\end{equation} 
This condition corresponds to the requirement that the classical pulse 1 
had enough time to overtake the $n$-photon pulse 2.
Because the mode function $\psi_2$ is normalized we thus find
\begin{align}
  \langle {\cal E}_1 (v_1 t,t) \rangle 
  &=  \alpha \sqrt{\eta} \, \psi_1(0) 
  e^{  -{\rm i}  n \phi_1} ,
\label{eq:QNDresult2}\\
  \phi_i &\equiv
   \frac{\eta v_{i} {\cal V}_{i}}{\Delta v} \; .
\label{phiDef}\end{align}
We remark that $\phi_1 = \phi_2$ for unitary dynamics. 
Eq.~(\ref{eq:QNDresult2}) is precisely the result that one would obtain in a single-mode
treatment. Hence, for unitary dynamics a mismatch in the group velocities
of two pulses that allows one pulse to overtake the other
in a Kerr medium will result in a uniform phase shift for
QND measurements of the photon number.
In practice, overtaking will impose a
minimum requirement on the length of the Kerr medium and
SPM will lead to a distortion of the signal.

\subsection{Controlled-Phase Gate} \label{sec:cpgUnitary}

Our proposal to build a CPG extends previous designs to build quantum gates using
XPM \cite{Chuu95,rebic06} by ensuring a uniform XPM phase shift. 
In a controlled phase gate, two qubits with logical basis
states $|0 \rangle , |1 \rangle $ are manipulated in such a way
that the state acquires a phase shift $\beta$ only if both qubits are
in the logical state $|1 \rangle \otimes |0 \rangle $.
In other words, a CPG is a unitary
operator that maps the basis states
$|kl \rangle \equiv |k \rangle \otimes | l \rangle $ to \cite{footnote3}
\begin{equation} 
   U_\text{CPG} |kl \rangle 
   = e^{{\rm i}\, k(1-l) \,\beta} |kl \rangle 
  \; ,\; k,l =0,1\; .
\end{equation} 
\begin{figure}[t]
\centering
\begin{tabular}{cc}
\includegraphics[width=8cm]{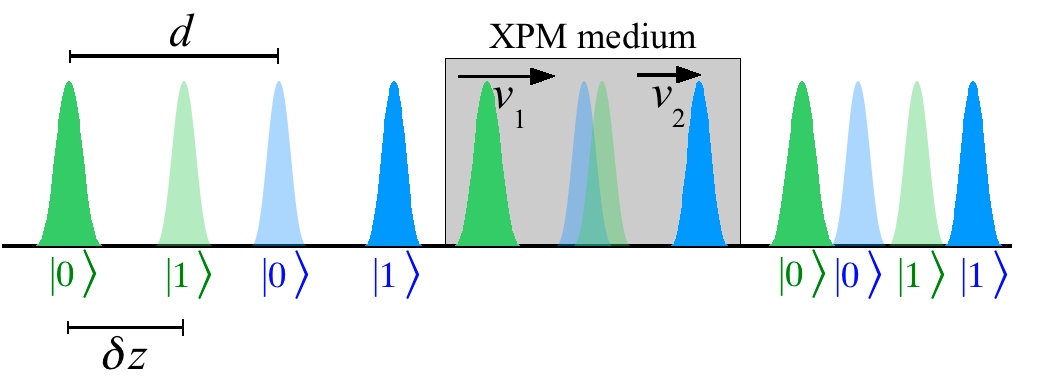}
\end{tabular}
\caption{\label{setup} 
Construction of a  controlled phase gate for two time-bin qubits using uniform XPM.
}
\end{figure}
With the uniform cross phase shifter discussed above, such a gate can
be implemented for time-bin encoded photonic qubits as indicated
in Fig.~\ref{setup}.
In this case each optical field ${\cal E}_n$ carries one photon.
For each photon the
logical basis states are encoded in two wavepackets, one
of which is delayed compared to the other. The logical state 
$|0 \rangle $ for qubit $n$ ($n= 1,2$) can be encoded in 
wavepackets $\psi_n(z)$ so that photons in this state are annihilated
by operator $a_n=a_n[0]$.
The logical state 
$|1 \rangle $ can be encoded in wavepackets that are shifted to the right,
$\psi_n(z-\delta z)$; photons in this state are annihilated by the operators
$a_n[\delta z]$.
A CPG can then be constructed by arranging the initial distances between the wavepackets in such
a way that the only states that do overlap inside the XPM medium are state $|1 \rangle $ of the first
qubit and state $|0 \rangle $ of the second qubit. For all other logical states the
wavepackets in the two field modes never overlap so that there is no XPM effect.

The logical basis states for two time-bin encoded qubits at time $t$ 
are given by the set
\begin{equation} 
  |ij(t) \rangle = a_1^\dagger[\delta z^{(1)}_i+v_1 t]\, a_2^\dagger[\delta z^{(2)}_j+v_2 t]\, |0 \rangle ,
\label{eq:logStates}\end{equation} 
for $(\delta z^{(1)}_i , \delta z^{(2)}_j) = (0,0), (0,\delta z), (\delta z,0),(\delta z ,\delta z)$
for $(i,j) = (0,0),(0,1),(1,0),(1,1)$, respectively. The factors $v_i t$ represent an explicit
time dependence in the definition of the logical states such that they are
co-moving with the wavepackets at group velocity $v_i$.

We consider the case that the two photons are initially prepared in a pure two-qubit
state $|\psi(0) \rangle =\sum_{kl} c_{kl} |kl(0) \rangle $. 
To characterize the performance of the CPG we need to obtain the
logical components $ \rho_{ij;kl}(t) = \langle ij(t)|\rho(t)|kl(t) \rangle $ of the density matrix $\rho(t)$ 
at time $t$. A somewhat tedious calculation that can be found in App.~\ref{app:rhoUnitaryDerivation}
leads to 
\begin{align} 
   \rho_{ij;kl}(t) &=    X_{kl}^* X_{ij} ,
\end{align}
with
\begin{align}
  X_{ij}&=
   c_{ij}  \int  dz_1\, dz_2\,  
    |\psi_1(z_1 ) \,
    \psi_2(z_2+d)|^2 
\nonumber \\ &\times
   e^{  -{\rm i}\eta \frac{ v_{2}}{\Delta v} 
  \theta_2 (z_2-z_1+d +\delta z^{(2)}_j-\delta z^{(1)}_i , \Delta vt)}.
\label{Xij2a}\end{align}
The phase factor in Eq.~(\ref{Xij2a}) determines the XPM effect. We now
prove that it is only nontrivial for the logical state $|10 \rangle $ because only in this
case the associated wavepackets overtake each other. 

The phase factor is proportional to the
function $ \theta_2 (z_2-z_1+d +\delta z^{(2)}_j-\delta z^{(1)}_i , \Delta vt)$
which, according to Appendix \ref{app:phaseShift}, is nearly constant for a given range
of its two variables. Ignoring the variables $z_i$ for the moment we consider
$ \theta_2 (d +\delta z^{(2)}_j-\delta z^{(1)}_i , \Delta v t)$ instead.
For the four states $00,01,10,11$ the first argument takes the
values $d, d+\delta z, d - \delta z$, and $d$ respectively. 
If we choose the time $t$ for which
the photons are interacting such that $d - \delta z < \Delta v t < d, d+\delta z$
then $\theta_2 =0$ for $ij = 00,01,11$ and $\theta_2 = {\cal V}_2 $ for $ij=10$.
Hence only the latter combination will experience an XPM phase shift. This conclusion
also holds if we re-introduce the variables $z_i$ because they only vary over the support 
of the two wavepackets, which is smaller than $d, \delta z$; the value of
$\theta_2$ therefore does not change. Thus we arrive at
\begin{align}
  X_{ij}&=
   c_{ij}  
     e^{  -{\rm i}  i(1-j) \phi_2},
\label{eq:Xij3}\end{align}
and the density matrix elements become
\begin{align} 
   \rho_{ij;kl}(t) &= 
   c_{ij} c_{kl}^* 
    e^{  -{\rm i}  (i(1-j)-k(1-l)) \phi_2} .
\label{eq:rhot}\end{align}

Let us now work out the performance of the CPG by calculating the
concurrence \cite{PhysRevLett.80.2245}
of the density matrix (\ref{eq:rhot}) for a specific initial product state $|\psi(0) \rangle =
(|0 \rangle + |1 \rangle )\otimes (|0 \rangle + |1 \rangle )/2$, which
corresponds to $c_{ij} = \frac{ 1}{2}$ for all $i,j$. 
The concurrence $C$ of the density matrix (\ref{eq:rhot}) then evaluates to $C = |\sin(\phi_2/2)|$.
This indicates a perfect CPG because the initial product state is transformed
into a maximally entangled state provided one can achieve a large phase shift
$\phi_2 = \pi$. 

\section{XPM with Non-Hamiltonian coupling} \label{sec:nonUnitary}

In the previous section we have devised a scheme to create uniform XPM for quantized light pulses
for the special case of unitary dynamics for which $v_1 V_1(z) = v_2 V_2(z)$. 
The theory behind this scheme corresponds to a regularized quantization of the classical 
theory (\ref{xpmEqnClass}) for the case that the XPM coefficients are related by $v_1 \gamma_1 = v_2\gamma_2$. 
However, in classical optics this relation between the two XPM coefficients is not generally adopted. 
In this section we therefore study the quantization of the classical model 
with $v_1 \gamma_1 \neq v_2\gamma_2$ and its implications for quantum information processing.
Because the model needs to be regularized, we consider the general case with
two non-local interaction potentials that fulfill $v_1 V_1(z) \neq v_2 V_2(z)$. 

The discussion in Sec.~\ref{Sec:pulseProp} has revealed that the quantum dynamics in this case must
be non-unitary because there is no Hamiltonian that can generate the equations of motion.
As Eq.~(\ref{xpmEqn}) is based on a phenomenological model that does not explicitly include
absorption or dissipation, the origin for this non-unitarity cannot be identified in an unambiguous
way. There might be implicit absorption processes hidden in the model, although we strongly doubt that this 
is the case because the dynamics does not have a structure that is comparable to the Lindblad form \cite{lindblad76}.
Another possibility is that the external fields that usually are needed for EIT-based XPM media
may induce energy fluctuations that lead to non-Hamiltonian dynamics.
Yet another possibility is that the direct quantization of the classical 
theory (\ref{xpmEqnClass}) ignores the averaging processes that are associated with a macroscopic
decription of electrodynamics \cite{jackson99}. Such averaging procedures lead to loss
of information, which may induce decoherence in a quantum system.

Despite this ambiguity with respect to the cause of non-unitary dynamics it is worthwhile
studying this situation because one can gain a better understanding of the quantization 
of a new class of classical models. Furthermore, even if the theory that we develop below
cannot include the microscopic details behind the XPM interaction it nevertheless
should help to estimate the effects of a mismatch between the XPM coefficients.

In the case  $v_1 V_1(z) \neq v_2 V_2(z)$,  Eq.~(\ref{solu}) still provides an exact solution 
to the dynamical equation of motion. However,
this solution is inconsistent with basic
requirements of quantum field theory. To illustrate this
point we consider the equal-time commutation relations between the
Heisenberg field operators, which should
agree with the commutation relations between the respective
Schr\"odinger operators. In our case this principle is violated
because
\begin{align} 
  {\cal E}_2(z',t) {\cal E}_1(z,t) &=  {\cal E}_1(z,t)   {\cal E}_2(z',t)   e^{-{\rm i} \Phi(z'-z,t)} ,
\\ 
   \Phi(z'-z,t) &\equiv \frac{ \eta}{\Delta v} \big (
   v_1  \theta_1(z'-z+\Delta v t,\Delta v t) 
\nonumber \\ & \hspace{7mm} - v_2  \theta_2(z'-z+\Delta v t,\Delta v t)  \big ),
\end{align} 
despite $[ {\cal E}_1(z,0) ,  {\cal E}_2(z',0)]=0$. 

The conventional way to deal with
non-Hamiltonian dynamics in the Heisenberg picture is to introduce
Langevin noise operators. In the context of nonlinear optics this has
first been done by Boivin {\em et al.} \cite{haus94} 
in the case of time-dependent self-phase modulation.

We now show that a similar approach can also be made for
instantaneous XPM between two light pulses with different group velocities.
Following Ref.~\cite{haus94} we introduce Hermitian decoherence operators
$m_i(z,t)$ such that the dynamical equations are modified to 
\begin{align}
  \left (\frac{1}{v_{i}}\partial_t+\partial_z \right ) {\cal E}_i(z,t)&=
-{\rm i} \Big ( \int_{-\infty}^{\infty} dz'I_{3-i}(z',t)
\label{xpmEqn2}\\ & \times 
V_i(z-z') + m_i (z,t)
   \Big )  {\cal E}_i(z,t) \; .
\nonumber \end{align}
The decoherence operators commute with all field operators. 
We adopt the frequently used assumption
of $\delta$-correlated decoherence,
\begin{equation} 
  [ m_i(z,t) \, , \, m_j(z', t') ] = {\rm i}\eta \left ( 
  \textstyle{ \frac{V_i(z-z')}{v_j}- \frac{ V_j(z-z')}{v_i} } \right )\delta(t-t') .
\label{noiseCom}\end{equation} 
The solution of Eq.~(\ref{xpmEqn2}) can be written as
\begin{align}
   {\cal E}_i(z,t) &= R_i(z,t) \, U_{3-i}(z,t) \, {\cal E}_i(z-v_i t,0),
\label{sol2}\\
  R_i(z,t) &= \exp \left (
   -{\rm i} v_i \int_0^t dt' \, m_i(z-v_i(t-t'), t')
  \right ) \; .
\end{align}
Using the commutation relations (\ref{noiseCom}) it is not hard to show that
\begin{align} 
  R_1^\dagger (z,t)\, R_2(z',t)\, R_1(z,t) &= 
   R_2(z',t)\,  e^{{\rm i}\Phi(z'-z,t)} ,
\nonumber \\
    R_2^\dagger (z,t)\, R_1(z',t)\, R_2(z,t) &= 
   R_1(z',t)\,  e^{-{\rm i}\Phi(z-z',t)}
\; .
\label{noiseTrans}\end{align}  
As a consequence, solution (\ref{sol2}) fulfills the equal-time
commutation relations. Introducing the decoherence term in 
Eq.~(\ref{xpmEqn2}) has thus led to a consistent quantum
field theory of XPM for light pulses with different group
velocities.

\subsection{Controlled-Phase Gate}

To characterize the performance of a CPG in the case of non-unitary dynamics we can repeat
the steps discussed in Sec.~\ref{sec:cpgUnitary} with solution (\ref{solu}) replaced by Eq.~(\ref{sol2}).
Finding the logical density matrix elements must then be done within
an open system approach that is described in App.~\ref{app:rhoNonUnitary}.
As a result, the factors $X_{ij} $ become operators on the Hilbert space of the environment
(on which the decoherence operators act) and Eq.~(\ref{eq:Xij3}) is replaced by
\begin{align}
  X_{ij}&=
   c_{ij} 
      e^{  -{\rm i}  i(1-j) \phi_2}
   \int  dz_1\, dz_2\,  
    |\psi_1(z_1 ) \,
    \psi_2(z_2+d)|^2 
\nonumber \\ &\times
   R_1(z_1+v_1 t+\delta z^{(1)}_i,t) 
  R_2(z_2+d +v_2 t+\delta z^{(2)}_j,t).
\label{Xij2A}\end{align}
The  logical density matrix elements take the form
\begin{align} 
   \rho_{ij;kl}(t) &= \text{Tr}_\text{E}  \big (\rho_\text{E}(0)
   X_{kl}^\dagger X_{ij} \big ),
\label{eq:rhoXXnoise}\end{align}
with $ \text{Tr}_\text{E}$ indicating the trace over the environment degrees of freedom and
$\rho_\text{E}(0)$ the initial state of the environment.

To keep the discussion concise we
assume that the photon pulses are very sharp. We then can 
replace the squares of the wavefunctions by $\delta$ distributions and get
\begin{align}
  X_{ij}&=
   c_{ij}  
     e^{  -{\rm i}  i(1-j) \phi_2}
   R_1(v_1 t+\delta z^{(1)}_i,t) 
R_2(d +v_2 t+\delta z^{(2)}_j,t) .
\label{eq:Xij3A}\end{align}
The density matrix elements then become
\begin{align} 
   \rho_{ij;kl}(t) &= 
   c_{ij} c_{kl}^* 
    e^{  -{\rm i}  (i(1-j)-k(1-l)) \phi_2}  N_{ij;kl}(t),
\label{eq:rhotA}\end{align}
with the Hermitian decoherence matrix
\begin{align}
   N_{ij;kl}(t) &\equiv
 \Big  \langle 
   R_2^\dagger (d +v_2 t+\delta z^{(2)}_l,t) 
   R_1^\dagger (v_1 t+\delta z^{(1)}_k,t)
 \nonumber \\ &\times
    R_1(v_1 t+\delta z^{(1)}_i,t)   
  R_2(d +v_2 t+\delta z^{(2)}_j,t) \Big \rangle  .
\end{align}
In App.~\ref{app:Nijkl}) we show that it can be written as
\begin{align}
  N &= \left ( \begin{array}{cccc}
           1 & c_2&  c_1   e^{{\rm i} (\phi_1-\phi_2)} &  N_{00;11} 
     \\ c_2^*  & 1 & N_{01;10} & c_1
     \\  c_1^* e^{-{\rm i} (\phi_1-\phi_2)} &  N_{01;10} ^* & 1 & c_2  
     \\ N_{00;11}^* & c_1^* &  c_2^*   & 1
     \end{array} \right ) ,
\label{noiseMat} \\
 c_n &\equiv  \Big  \langle 
  R_n^\dagger (v_n t+\delta z,t) R_1(v_n t,t) \Big \rangle  .
\end{align}
Hence the influence of decoherence in our model can be described
by four complex parameters $N_{00;11} $, $N_{01;10}$, and $c_n$.
This is a consequence of the approximation of very short pulses.
For extended pulses additional decoherence contributions are anticipated.

We again calculate the concurrence
of density matrix (\ref{eq:rhotA}) for the initial product state
$c_{ij} = \frac{ 1}{2}$ for all $i,j$.
The specific values of the parameters
$N_{00;11} $, $N_{01;10}$, and $c_n$ depend on the detailed decoherence model that
one employs so that the predictions may vary substantially
depending on the assumptions behind the model. We present
one particular decoherence model in Appendix \ref{app:noiseModel}. However, 
all decoherence models must be consistent with all eigenvalues of the density matrix  (\ref{eq:rhot}) 
taking values between 0 and 1. 

\begin{figure}[t]
\includegraphics[width=6cm]{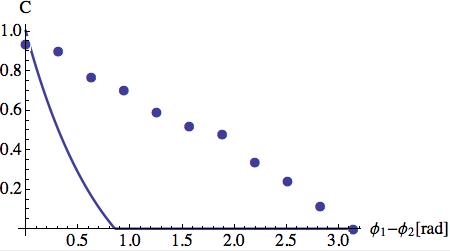} 
\caption{\label{fig:concurrence} 
Concurrence as a function
of the XPM phase shift for minimal decoherence (dots) and the decoherence
model presented in Appendix \ref{app:noiseModel} (solid line).
}
\end{figure}

To simplify the discussion we consider the special case  $\phi_2 = \pi$ for the oft-used assumption
that the two interaction potentials $V_i(z)$ are equal and independent of the group velocities
\cite{PTL6-733,JRLR29-57} so that ${\cal V}_1 = {\cal V}_2$.
The phase factor that appears in the decoherence matrix (\ref{noiseMat})
is then $\phi_1 -\phi_2 = \pi \Delta v /v_2 >0$.
For a given value of $\phi_1 -\phi_2$ we can derive a ``minimal decoherence model''
by finding those values of the parameters
$N_{00;11} $, $N_{01;10}$, and $c_n$ which maximize the concurrence $C$
for a consistent density matrix.
Fig.~\ref{fig:concurrence} shows the concurrence for this minimal decoherence model
as a function of  $\phi_1 -\phi_2$. The dots are found through a random search 
for the parameters $N_{00;11} $, $N_{01;10}$, and $c_n$ of the minimal decoherence model. 
Each dot is based on a sample of typically $10^7$ random events. The error in each
value is estimated to be about 5\%.
The concurrence decreases
with  $\phi_1 -\phi_2 $ and becomes zero for  $\phi_1 -\phi_2 =\pi$. In this
case decoherence completely destroys the entangling capacity of the XPM
interaction. The numerically determined density matrix then corresponds to a nearly
equal mixture of two highly entangled states.
Realistic decoherence models would typically
predict a less than optimal performance.
For instance, the blue line in Fig.~\ref{fig:concurrence} shows $C$ for the 
decoherence model of App.~\ref{app:noiseModel} under the fairly optimistic
assumption $\Delta v t = 4 \sqrt{\pi} w$, where $w$ is the width of
a Gaussian potential $V_1(z)=V_2(z)= V(0) \exp(-z^2/w^2)$. For more realistic values
$\Delta v t \gg w$ \cite{footnote4}
the concurrence would be non-zero only in a 
very narrow range around $\phi_1 = \phi_2$.

Our phenomenological model therefore
suggests that a CPG  may not be achievable with XPM unless 
$\phi_1 $ is very close to $\phi_2$. In most proposals for XPM based on
double EIT this could be achieved by a suitable preparation of the atomic
gas, although it may require fine tuning of parameters like magnetic fields
or pump field intensities. A better way to achieve $\phi_1=\phi_2$ would
be to find a system in which this is guaranteed through microscopic
symmetries.

\subsection{Quantum non-demolition measurement of photon numbers}
The scheme for a QND measurement of the photon number discussed
in Sec.~\ref{sec:qndUnitary} can easily be extended to the non-unitary
case by calculating the measurement signal
$\langle  {\cal E}_1 (z,t)  \rangle $
using solution (\ref{sol2}) instead of (\ref{solu}). The only change is that the measurement result
(\ref{eq:QNDresult2}) is multiplied by $ \langle R_1(z,t) \rangle $. This factor depends
strongly on the specific decoherence model but generally will lead to a decrease in the
contrast of the phase measurement.

To give a rough estimate we consider the decoherence model
of App.~\ref{app:noiseModel} in which the noise is generated by an environment consisting
of a harmonic oscillator field with a flat frequency distribution. We then have 
\begin{equation} 
  \langle R_i(z,t) \rangle =
  \exp \left (
    - \frac{ \eta}{4} \frac{ v_i}{v_{3-i}} \Delta v \, t \, V(0)
  \right ).
\label{eq:meanR1}\end{equation} 
The decoherence amplitude $ \langle R_i(z,t) \rangle $ can vary between the optimal
value 1, if the interaction potential $V(z) = V_1(z)=V_2(z)$  vanishes at the origin, and very small values for
a sharply peaked potential. We remark that this result also demonstrates the necessity of a 
non-local potential to avoid the diverging quantum fluctuations in the case
$V(z)\propto \delta(z)$ \cite{shapiro:josab1993}.

It is instructive to evaluate this for a Gaussian potential
$V(z) = V(0) \exp(-z^2/w^2)$ where $V(0)>0$ and the width $w$ should be
not substantially larger than an atom. We then find
$\phi_1-\phi_2 \approx \sqrt{\pi}V(0) w/\eta$ so that
\begin{equation} 
  \langle R_i(z,t) \rangle \approx 
  \exp \left (
    - \frac{\phi_1-\phi_2 }{4\sqrt{\pi}} \frac{ v_i}{v_{3-i}} \frac{ \Delta v \, t}{w} 
  \right ).
\label{eq:meanR2} \end{equation} 
The first factor is proportional to the XPM phase shift difference.
The last factor $\Delta v\, t/w$ is typically much larger than one
because the length $\Delta v\, t$ must be larger than the
width of the light pulses ($t$ is the time needed for one pulse
to overtake the other). Hence, for very short (femtosecond)
pulses the exponential may be of the same size as the XPM phase
shift. Typically, however, it will be much larger.
We therefore conjecture that for the QND measurement
to be successful the XPM phase shifts $\phi_i$ in the two light pulses must also
be nearly equal.

\section{ Conclusion}
In this paper we have devised a scheme to generate a uniform XPM for quantum
information processing and applied it to characterize the fidelity of a QND
measurement of the number of photons in a light pulse and the performance
of a deterministic CPG. The analysis is based on a regularized quantization
(\ref{xpmEqn}) of a common phenomenological model for
the classical XPM effect between two light pulses with different group velocities.

When the pulses overtake each other while traveling through a nonlinear medium
the induced XPM phase shift $\phi_i, i=1,2,$ is uniform over each pulse.
For $\phi_1=\phi_2$ the dynamics is unitary and predicts perfect fidelities for
QND measurement and CPG. For $\phi_1\neq \phi_2$
a decoherence effect must be introduced to keep the model consistent, and this decoherence
significantly affects the fidelity of the gate and the QND measurement.

 \section*{Acknowledgments} 
We thank A.~I.~Lvovsky for
helpful discussions. This work has been supported
by NSERC, iCORE, and QuantumWorks. 
B.~C.~S.~is a CIFAR Associate.
S.~A.~M.~acknowledges support from
the Russian Foundation of Basic Research under grant \# 08-07-00449.

\begin{appendix}
\section{Useful commutation relations}
In this appendix we will summarize a number of commutation relations that will
be useful in deriving our main results.

The effect of XPM on the pulse propagation is governed by the action of the
unitary operators $U_i= U_i (z,t)$ of Eq.~(\ref{eq:U1}) on the field operators,
\begin{align}
   U_{i}^\dagger {\cal E}_i(z',0) U_{i} &= 
    {\cal E}_i(z',0)
\exp \Big (
  -{\rm i}\eta \frac{ v_{3-i}}{v_i - v_{3-i}} 
\nonumber \\ &\times
\theta_{3-i} (z-z'-v_{3-i} t , (v_i-v_{3-i})t)
  \Big ),
\label{unitrans}\\
   U_{i} {\cal E}_i(z',0) U_{i}^\dagger &= 
    {\cal E}_i(z',0)
\exp \Big (
  {\rm i}\eta \frac{ v_{3-i}}{v_i - v_{3-i}} 
\nonumber \\ &\times
\theta_{3-i} (z-z'-v_{3-i} t , (v_i-v_{3-i})t)
  \Big ).
\label{unitrans2}
\end{align}

The effect of XPM on the annihilation operator for photons
in mode $i$ can be derived from the commutation relations
\begin{align}
  [a_i[u] , \mathcal{E}_i^{\dagger}(z,0)]&= \sqrt{\eta} \psi_i^{*}(z-u),
\label{eq:comm1}\end{align}
and
\begin{equation} 
  U_i^\dagger (z,t) \, a_i[u] U_i(z,t) = \tilde{a}_i[u; z,t],
\label{eq:unitransA}\end{equation} 
with $\tilde{a}_i[u; z,t]$  of Eq.~(\ref{eq:aTilde}).

\section{Simplification of the phase shift}\label{app:phaseShift}
\begin{figure}[t]
\includegraphics[width=4cm]{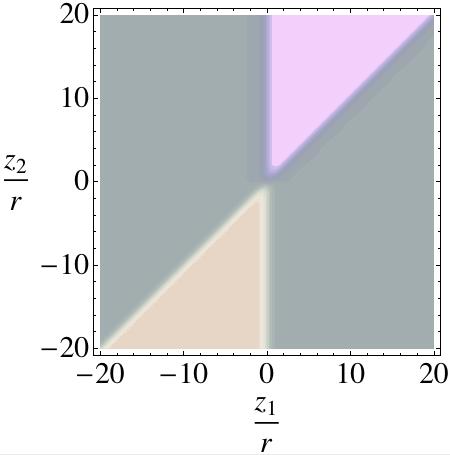}
\caption{\label{ThetaR} 
The function $\theta_i(z_1,z_2)$ for a Gaussian potential $V_i(z)$ of width $r$.
}
\end{figure}

If the potential $V_i(z)$ has a finite range $r$, so that
it vanishes for $|z|>r$, then $\theta_i(z_1,z_2)$ has generally a shape
similar to Fig.~\ref{ThetaR}.
It is essentially zero outside
the ``edges'' depicted in the figure and has an extended plateau
between the lines $z_1=0$ and $z_1=z_2$. 
The value on the plateau is given by
$\text{sign}(z_2)  {\cal V}_i$, with
${\cal V}_i$ defined in Eq.~(\ref{calViDef}).
The only region where $\theta_i(z_1,z_2)$ varies
significantly are bands of width $2r$ around the lines  
$z_1=0$ and $z_1=z_2$, and only
this transition region depends on the details of the potential.
If the range $r$ is much smaller than the optical wavelength 
we can ignore the transition region so that
$\theta_i(z_1,z_2)$ can be replaced by the plateau value between the
lines,
\begin{equation} 
  \theta_i(z_1, z_2) \approx
\left (
  \theta (z_1) \, \theta (z_2 -z_1) -  \theta (-z_1) \, \theta (z_1 -z_2)
  \right ) {\cal V}_i\; ,
\end{equation} 
with $\theta (z)$ the step function.

\section{Derivation of the action of a CPG on the density matrix}

\subsection{Unitary dynamics}\label{app:rhoUnitaryDerivation}
The evolution of the density matrix $\rho$ can be expressed with the aid of a
unitary operator $U(t)$ such that $\rho(t) = U(t) \rho(0) U^\dagger (t)$, with
\begin{equation} 
  \rho(0) = \sum_{i',j',k,'l'} c_{k'l'}^* c_{i'j'} |i'j'(0) \rangle \langle k'l'(0) | .
\label{initRho} \end{equation} 
The logical matrix elements of the density matrix can then be expressed
as $\rho_{ij;kl}(t) = \langle ij(t) | U(t) \rho(0) U^\dagger (t) |kl(t) \rangle $.
Using the definition (\ref{eq:logStates}) of the logical basis states
we are then led to consider the expression
\begin{align} 
  \langle ij(t)|U(t) &= \langle 0| U^\dagger(t) a_1[\delta z^{(1)}_i+v_1 t]\, 
  a_2[\delta z^{(2)}_j+v_2 t]\, U(t) .
\label{eq:step1}\end{align}
We here inserted an additional factor $U^\dagger (t)$ which does not change the result
because $U(t) |0 \rangle = |0 \rangle $.
The operators $a_i[z]$ are defined using the Schr\"odinger operator ${\cal E}_i(z,0)$,
see Eq.~(\ref{eq:aModeOp}). The effect of $U^\dagger \cdots U(t)$ is now to replace the Schr\"odinger
operator by the respective Heisenberg operator ${\cal E}_i(z,t)$ so that
\begin{align} 
  \langle ij(t)|U(t) &= \langle 0| a_1[\delta z^{(1)}_i+v_1 t,t]\, 
  a_2[\delta z^{(2)}_j+v_2 t,t],
\label{eq:step2} \\
   a_i[z,t] & \equiv \frac{1}{\sqrt{\eta}}\int_{-\infty}^{\infty} dz\,
\psi^{*}_i(z-u)\, \hat {\cal{E}}_i(z,t) \; .
\end{align}
We thus find for the logical matrix elements
\begin{align} 
   \rho_{ij;kl}(t) &= 
  \langle \psi(0) |   a_2^\dagger [\delta z^{(2)}_l+v_2 t,t]
  a_1^\dagger [\delta z^{(1)}_k+v_1 t,t]\, 
    |0 \rangle 
  \nonumber \\ &\times
  \langle 0|
   a_1[\delta z^{(1)}_i+v_1 t,t]\, 
   a_2[\delta z^{(2)}_j+v_2 t,t] |\psi (0) \rangle .
\label{eq:step3}\end{align}
We now concentrate on the factor
\begin{align}
  X_{ij}&\equiv\langle 0|
   a_1[\delta z^{(1)}_i+v_1 t,t]\, 
   a_2[\delta z^{(2)}_j+v_2 t,t] |\psi (0) \rangle
\nonumber \\ &\hspace{-3mm}= 
   \frac{1}{\eta}\int  dz_1\, dz_2\,
    \psi^{*}_1(z_1-v_1 t-\delta z^{(1)}_i)\, 
\nonumber \\ &\times
    \psi^{*}_2(z_2-v_2 t-\delta z^{(2)}_j)\, 
    \langle 0| 
    U_2(z_1,t)  \hat {\cal{E}}_1(z_1-v_1t,0) 
\nonumber \\ &\times
  U_1 (z_2,t) \hat {\cal{E}}_2(z_2-v_2 t,0)
   |\psi(0) \rangle 
\nonumber \\ &\hspace{-3mm}= 
   \frac{1}{\eta}\int  dz_1\, dz_2\,  
    \psi^{*}_1(z_1-v_1 t-\delta z^{(1)}_i)\, 
\nonumber \\ &\times
    \psi^{*}_2(z_2-v_2 t-\delta z^{(2)}_j)\, 
    \langle 0| 
   U_1^\dagger  (z_2,t)  \hat {\cal{E}}_1(z_1-v_1t,0) 
\nonumber \\ &\times
   U_1 (z_2,t) \hat {\cal{E}}_2(z_2-v_2 t,0)
   |\psi(0) \rangle 
\nonumber \\ &\hspace{-3mm}= 
   \frac{1}{\eta}\int  dz'_1\, dz'_2\,  
    \psi^{*}_1(z'_1 )\, 
    \psi^{*}_2(z'_2+d)
\nonumber \\ &\times
   e^{  -{\rm i}\eta \frac{ v_{2}}{\Delta v} 
  \theta_2 (z'_2-z'_1+d +\delta z^{(2)}_j-\delta z^{(1)}_i , \Delta vt)}
\nonumber \\ &\times
    \langle 0| 
  \hat {\cal{E}}_1(z'_1+\delta z^{(1)}_i ,0) 
  \hat {\cal{E}}_2(z'_2+d+\delta z^{(2)}_j ,0)
   |\psi(0) \rangle .
\label{Xij1}\end{align}
The expectation value in the last line can be considerably simplified.
\begin{align}
    \langle 0| 
  \hat {\cal{E}}_1&(z'_1+\delta z^{(1)}_i ,0) 
  \hat {\cal{E}}_2(z'_2+d+\delta z^{(2)}_j ,0)
   |\psi(0) \rangle 
\nonumber \\ &\hspace{-3mm}=
  \sum_{k,l} c_{kl}
      \langle 0| 
  \hat {\cal{E}}_1(z'_1+\delta z^{(1)}_i ,0) 
  \hat {\cal{E}}_2(z'_2+d+\delta z^{(2)}_j ,0)
\nonumber \\ & \times
  a_1^\dagger [\delta z^{(1)}_k] a_2^\dagger [\delta z^{(2)}_l] 
   |0\rangle 
\nonumber \\ &\hspace{-3mm}=
  \eta \sum_{k,l} c_{kl}  \psi_1(z'_1+\delta z^{(1)}_i -\delta z^{(1)}_k )
\nonumber \\ & \times
  \psi_2(z'_2+d+\delta z^{(2)}_j - \delta z^{(2)}_l).
\label{ijklFactor}\end{align}
The wavefunctions in Eq.~(\ref{Xij1}) imply that $z'_i \ll \delta z$
because $z'_i$ can only vary over the support of the wavefunctions.
This in turn implies in Eq.~(\ref{ijklFactor}) that
$\delta z^{(1)}_i =\delta z^{(1)}_k $ and 
$\delta z^{(2)}_j = \delta z^{(2)}_l$ because otherwise
the argument of the wavefunctions would be outside of
their support. Hence,
\begin{align}
    \langle 0| 
  \hat {\cal{E}}_1(z'_1+\delta z^{(1)}_i ,0) 
  \hat {\cal{E}}_2 &(z'_2+d+\delta z^{(2)}_j ,0)
   |\psi(0) \rangle 
\nonumber \\ &=
  \eta c_{ij}  \psi_1(z'_1 )
 \psi_2(z'_2+d),
\label{ijklFactor2}\end{align}
from which Eq.~(\ref{Xij2a}) follows.

\subsection{Non-Hamiltonian Dynamics}\label{app:rhoNonUnitary}

The derivation of the logical density matrix elements for non-unitary evolution
follows essentially the same steps as in the unitary case described
in App.~\ref{app:rhoUnitaryDerivation}. However, because the decoherence operators
act on a different Hilbert space ${\cal H}_\text{E}$ associated with the environment, 
the calculations have to be made on an enlarged Hilbert
space ${\cal H}_\text{S+E} = {\cal H}_\text{S} \otimes {\cal H}_\text{E} $ that includes
our system of photon pulses as well as the environment. The dynamics on this enlarged
Hilbert space is described by a unitary operator $U(t)$ and we assume that 
initially system and environment are uncorrelated, 
$\rho_\text{S+E}(0) = \rho(0) \otimes \rho_\text{E}(0)$, with $\rho(0)$ of Eq.~(\ref{initRho}).
The logical matrix elements are then given by
\begin{align} 
   \rho_{ij;kl}(t) &= \text{Tr}_\text{E}  \big (
    \langle ij(t)|U(t)  \rho(0) \otimes \rho_\text{E}(0) U^\dagger (t) |kl(t) \rangle \big ),
\label{rho1A}\end{align}
with $ \text{Tr}_\text{E} $ the trace over the environment.

We again start the derivation with term (\ref{eq:step1}), but now
$U$ refers to the unitary evolution on the total Hilbert space.
If $U$ leaves the ground state invariant, in the sense that
 $U|0 \rangle \otimes |E \rangle =  |0 \rangle \otimes U_\text{E} |E \rangle$ 
for any state $|E \rangle $ of the environment and a fixed
unitary map $U_\text{E}$ that acts on the environment, then  
we can again make the transition to Heisenberg operators as in
Eq.~(\ref{eq:step2}). The logical matrix elements
then can be written in the form (\ref{eq:rhoXXnoise}) with the
environment operators 
\begin{align}
  X_{ij}&\equiv   \langle 0|
   a_1[\delta z^{(1)}_i+v_1 t,t]\, 
    a_2[\delta z^{(2)}_j+v_2 t,t] |\psi (0) \rangle
\nonumber \\ &= 
  \frac{1}{\eta}\int  dz_1\, dz_2\,  R_1(z_1,t) R_2(z_2,t)
    \psi^{*}_1(z_1-v_1 t-\delta z^{(1)}_i)\, 
\nonumber \\ &\times
    \psi^{*}_2(z_2-v_2 t-\delta z^{(2)}_j)\, 
e^{  -{\rm i}\eta \frac{ v_{2}}{\Delta v} 
\theta_2 (z_2-z_1+\Delta v t , \Delta vt)}
\nonumber \\ &\times
    \langle 0| 
  \hat {\cal{E}}_1(z_1-v_1t,0) 
  \hat {\cal{E}}_2(z_2-v_2 t,0)
   |\psi(0) \rangle .
\label{Xij1A}\end{align}
This expression corresponds to Eq.~(\ref{Xij1}). The argument that leads to
Eq.~(\ref{ijklFactor2}) can be repeated, resulting in
\begin{align}
  X_{ij}&=
   c_{ij}  \int  dz_1\, dz_2\,  
    |\psi_1(z_1 ) \,
    \psi_2(z_2+d)|^2 
\nonumber \\ &\times
   R_1(z_1+v_1 t+\delta z^{(1)}_i,t) 
  R_2(z_2+d +v_2 t+\delta z^{(2)}_j,t)
\nonumber \\ &\times
   e^{  -{\rm i}\eta \frac{ v_{2}}{\Delta v} 
  \theta_2 (z_2-z_1+d +\delta z^{(2)}_j-\delta z^{(1)}_i , \Delta vt)},
\label{Xij2aA}\end{align}
which corresponds to Eq.~(\ref{Xij2a}). The discussion of the phase factor that
leads to Eq.~(\ref{eq:Xij3}) is unaffected by the decoherence terms so that
Eq.~(\ref{Xij2A}) can be deduced.

\subsection{Derivation of the decoherence matrix} \label{app:Nijkl}
The decoherence matrix can be evaluated without further
assumptions in a number of cases.
Because $R_i^\dagger(z,t) R_i(z,t) =1$ it is easy to see that
the diagonal elements are equal to one. This also guarantees 
trace preservation.
For the same reason,  whenever $i=k$ the decoherence operators simplify to
\begin{align} 
    N_{ij;il}(t) &=
 \Big  \langle 
   R_2^\dagger (d +v_2 t+\delta z^{(2)}_l,t) 
  R_2(d +v_2 t+\delta z^{(2)}_j,t) \Big \rangle  .
\end{align} 
Furthermore, for $l=j$ $R_2^\dagger $ and $R_2$ have the same argument so that
we can make use of Eq.~(\ref{noiseTrans}):
\begin{align}
   N_{ij;kj}(t) &=
 \Big  \langle 
   R_2^\dagger (d +v_2 t+\delta z^{(2)}_j,t) 
   R_1^\dagger (v_1 t+\delta z^{(1)}_k,t)
 \nonumber \\ &\times
    R_1(v_1 t+\delta z^{(1)}_i,t) 
  R_2(d +v_2 t+\delta z^{(2)}_j,t) \Big \rangle  
 \nonumber \\ &\hspace{-3mm}=
    \Big  \langle 
  R_1^\dagger (v_1 t+\delta z^{(1)}_k,t)
   R_1(v_1 t+\delta z^{(1)}_i,t) \Big \rangle  
 \nonumber \\ &\times
 e^{{\rm i}\Phi(d -\Delta v t +\delta z^{(2)}_j-\delta z^{(1)}_k,t) 
  - {\rm i}\Phi(d -\Delta v t+\delta z^{(2)}_j-\delta z^{(1)}_i, t) }
\nonumber \\ &\hspace{-3mm}= 
     \Big  \langle 
  R_1^\dagger (v_1 t+\delta z^{(1)}_k,t)
   R_1(v_1 t+\delta z^{(1)}_i,t) \Big \rangle  
\nonumber \\ &\times
  e^{{\rm i} (\phi_1-\phi_2)(1-j)(k-i) }.
\end{align}

\section{A specific decoherence model} \label{app:noiseModel}
The following model is inspired by the decoherence model for
time-dependent Kerr nonlinearities of Ref.~\cite{Shap06}
and restricted to the case $V_1(z) = V_2(z)\equiv V(z)$.
The reservoir that produces the decoherence is composed out
of harmonic oscillators in their ground state.
The decoherence operators are given by
\begin{align}
  m_1(z,t) &= M(z,t) + M^\dagger (z,t),
\\
  m_2(z,t) &= -{\rm i} M(z,t) + {\rm i} M^\dagger (z,t),
\\
  M(z,t) &= \sqrt{\frac{ \eta\Delta v}{4\pi  v_1 v_2}} \int_0^\infty\!\!
  d\omega\, ({\rm i} B(z,\omega) e^{-{\rm i}\omega t}  + C (z,\omega)e^{{\rm i}\omega t} ),
\end{align}
with
\begin{align} 
  [B(z,\omega) , B^\dagger (z',\omega')] &= 
   [C(z,\omega) , C^\dagger (z',\omega')] 
\nonumber \\ &= V(z-z') \delta(\omega-\omega').
\end{align} 
The operators $m_i$ fulfill the commutation relations (\ref{noiseCom}).
Using the Baker-Campbell-Hausdorff equation to separate the
annihilation part $M(z,t)$ and the creation part $M^\dagger (z,t)$ in the
exponent Eq.~(\ref{eq:meanR1}) can be proven.
With the same method a straightforward but tedious calculation
yields
\begin{align} 
  c_i &= \langle R_i(z,t) \rangle^2,
\\
 N_{00;11} &=  N_{01,10} = e^{\frac{ {\rm i}}{2}(\phi_1-\phi_2)} c_1 c_2.
\end{align}

\end{appendix}


\begin{thebibliography}{00}
\bibitem{footnote1}
  Answering the question whether the case $v_1 V_1(z)\neq v_2 V_2(z)$
  of this phenomenological model describes a real physical system 
  may require an ab-initio quantum description
  of a medium that supports XPM. 
  This is a formidable task and beyond the aim of our work to propose schemes
  for QND measurements of the photon number and to generate a CPG.
\bibitem{footnote3}
  Usually the state in which the phase shift is acquired is taken to be $|11 \rangle $,
  but we consider $|10 \rangle $ to simplify the discussion.
  Both gates are related by a single-qubit NOT operation that acts on the second qubit.
\bibitem{footnote4}
  For instance, for giant nonlinearities based on EIT  we would have  
  $\Delta v t \gg w$
  because the light pulses would have a duration in the order of 
  $\mu$s.
\end{thebibliography}
\end{document}